\documentclass[trackchanges]{aastex701}
\usepackage{CJK}
\usepackage{multirow}
\usepackage{fontenc}
\usepackage[figuresright]{rotating}
\usepackage{tabularx}
\usepackage{makecell}
\usepackage{url}

\begin{document}
\begin{CJK*}{UTF8}{gbsn}
\title{An Enhanced Sample of Galactic Red Supergiants Reveals Spiral Structures}

\author[orcid=0000-0002-3828-9183]{Zehao Zhang (张泽浩)}
\affiliation{School of Physics and Astronomy, Beijing Normal University, Beijing 100875, China}
\affiliation{Institute for Frontiers in Astronomy and Astrophysics, Beijing Normal University, Beijing 102206, China}
\email{zhzhang@mail.bnu.edu.cn}

\author[orcid=0000-0003-3168-2617]{Biwei Jiang (姜碧沩)}
\affiliation{School of Physics and Astronomy, Beijing Normal University, Beijing 100875, China}
\affiliation{Institute for Frontiers in Astronomy and Astrophysics, Beijing Normal University, Beijing 102206, China}
\email{bjiang@bnu.edu.cn}

\author[orcid=0000-0003-1218-8699]{Yi Ren (任逸)}
\affiliation{College of Physics and Electronic Engineering, Qilu Normal University, Jinan 250200, China}
\email{yiren@qlnu.edu.cn}

\correspondingauthor{Biwei Jiang}
\email{bjiang@bnu.edu.cn}

\begin{abstract}

Red supergiants (RSGs), representing a kind of massive young stellar population, have rarely been used to probe the structure of the Milky Way, mainly due to the long-standing scarcity of Galactic RSG samples. The Gaia BP/RP spectra (hereafter XP), which cover a broad wavelength range, provide a powerful tool for identifying RSGs. In this work, we develop a feedforward neural network classifier that assigns to each XP spectrum a probability of being an RSG, denoted as $\mathrm{P(RSG)}$. We perform ten independent runs with randomly divided training and validation sets, and apply each run to all XP spectra of stars with $G < 12$ mag. By selecting sources with $\mathrm{P(RSG)} \geq 0.9$, ten high-confidence candidate samples are obtained. A star is considered a ture Galactic RSG only if it appears in at least eight of these samples, yielding a final catalog of 2,436 objects. These RSGs show a clear spatial correlation with OB stars and trace the Galactic spiral arms well, confirming the reliability of our classification, and highlighting their potential to serve as powerful tracers of the Milky Way's structure.

\end{abstract}



\section{Introduction}  \label{sec:intro}

Red supergiants (RSGs) are evolved massive stars in the core-helium-burning phase, with initial masses of about 8$-$40 $M_{\odot}$ \citep{1979ApJ...232..409H}. Investigating their physical properties is important for understanding the evolution of massive stars, and it is essential to construct RSG samples that are as complete and clean as possible. In recent years, the number of known RSGs in the Local Group has been significantly increased. Thousands of RSGs are identified in nearby galaxies such as the Small and Large Magellanic Clouds (SMC and LMC), Andromeda (M31), and Triangulum (M33) (e.g., \citealt{2021ApJ...907...18R, 2021ApJ...923..232R}). These extensive samples have directly advanced studies of RSG variability \citep{2024ApJ...969...81Z}, binary fraction \citep{2021ApJ...908...87N, 2022MNRAS.513.5847P, 2025MNRAS.539.1220D, 2025ApJ...988...60D}, mass loss \citep{2021ApJ...912..112W, 2023A&A...676A..84Y, 2024AJ....167...51W, 2024A&A...686A..88A, 2025A&A...702A.178A}, and star formation histories \citep{2024ApJ...966...25R}.

As cool and luminous young objects, RSGs are among the brightest stars in the near-infrared, making them excellent tracers of spiral structures of galaxies, which has been proven in M33 and M31 (see Figure 11 of \citealt{2021ApJ...907...18R}). In the Milky Way, however, spiral structures are typically described using tracers like OB stars \citep{2018RAA....18..146X, 2019MNRAS.487.1400C}, H II regions \citep{2015ApJ...800...53H, 2015AJ....150..147F}, molecular gas \citep{2016ApJ...822...52R, 2016PASJ...68....5N}, and 21\,cm line emission \citep{2007AJ....134.2252S, 2009ApJ...693.1250D}. The use of RSGs as spiral structure tracers in our Galaxy is hindered by the incompleteness of known RSG samples. Although \citet{1989IAUS..135..445G} predicted roughly 5,000 RSGs in the Milky Way, the number of published Galactic RSGs over the last two decades remains comparatively limited. For example, \citet{2005ApJ...628..973L} analyzed 74 Galactic RSGs via optical spectrophotometry with MARCS atmosphere modelling; \citet{2018MNRAS.475.2003D} built a catalog of 197 cool supergiants using a 2MASS photometric pre-selection followed by Ca II triplet spectroscopy; \citet{2019AJ....158...20M} and \citet{2023A&A...671A.148M, 2025A&A...698A.282M} compiled 762 RSG candidates using Gaia parameters and BP/RP spectra with near-IR (2MASS/WISE) constraints; and \citet{2024MNRAS.529.3630H} compiled 578 highly probable and 62 likely Galactic RSG candidates by combining Gaia DR3 distances, 2MASS photometry, and 3D dust maps. These catalogs show substantial overlap with each other. It should be noted that the use of 2MASS data has limited the full use of the Gaia database. Consequently, most previous studies of Galactic RSGs have focused on identifying sources within individual spiral arms \citep{2018MNRAS.475.2003D}, detecting Galactic tangent points to spiral arms with RSGs \citep{2025A&A...698A.282M}, or qualitatively examining their spatial correlation with OB stars \citep{2024MNRAS.529.3630H}. Indeed, the paucity of datapoints does not allow using them as global tracers of the Milky Way's spiral structure.

The primary obstacle for the small number of identified Galactic RSGs is the significant and inhomogeneous interstellar extinction across the Galactic disk, which makes accurate extinction correction difficult \citep{2003A&A...409..205D}. As a result, it is impractical to select RSGs using traditional photometric methods based on color-magnitude diagrams (CMD). More recently, our group has taken several new approaches to improve the identification of Galactic RSGs. Using Gaia DR3 RVS spectra, \citet{2025A&A...694A.152Z} identified $\sim$6,200 RSG candidates across the Milky Way, providing what is considered a more realistic estimate of the Galactic RSG population \citep{2025Galax..13...66B}. In addition, \citet{2025MNRAS.538..101Z} combined OGLE time-series photometry with Gaia DR3 stellar parameters to identify 474 RSGs on the other side of the Galaxy, based mainly on their irregular light variations dominated by granulations \citep{2020ApJ...898...24R, 2024ApJ...969...81Z}, which revealed the flare structure of the far side of the Milky Way. However, RSGs exhibit only subtle differences from asymptotic giant branch stars (AGBs) and luminous red giants in their RVS spectral features and variability properties, making contamination in an RVS-based sample unavoidable. This limitation is mainly caused by the narrow wavelength coverage of the Gaia RVS (846-870 nm), which is not able to make a net separation of RSGs and other evolved stars. The contamination rate is also difficult to quantify. The blue and red prism spectra (BP/RP, hereafter XP) released by Gaia, however, cover a much wider wavelength range, containing several characteristic spectral lines of RSGs, mainly TiO. These features make the XP spectra highly suitable for machine-learning classification of massive stars like RSGs (see, e.g., \citealt{2023ApJ...959..102D}). Moreover, XP spectra can well reflect the stellar parameters of stars \citep{2023A&A...674A..26C, 2023A&A...674A..28F}, with a release quantity of $\sim$220 million, it is possible to more accurately and comprehensively identify RSGs in the Milky Way. In this work, we apply a neural-network-based classifier to XP spectra to enlarge the sample of Galactic RSGs, and attempt to provide an independent tracer of the Milky Way's spiral structure (independent because maximising the identifications and using pure Gaia data, without combining it with limiting external datasets).

The paper is organized as follows: Section \ref{sec:data} introduces the XP spectra and the training sample, as well as the pre-processing steps. Section \ref{sec:model} describes the neural network model and the training process. Section \ref{sec:result} presents the results of the classification, provides the Galactic RSGs sample obtained in this work, and discusses the spatial distributions of the RSGs. Finally, Section \ref{sec:summary} concludes the paper.

\section{Data and Pre-processing} \label{sec:data}

\subsection{Gaia XP Spectra} \label{sec:XP_spectra}

Gaia DR3 provides low-resolution spectra, known as XP spectra, for approximately 220 million stars, with a spectral resolution of $R\sim20-100$ \citep{2023A&A...674A...1G}. The XP spectra cover a wavelength range of $330-1050$ nm, extending from optical to near-infrared, and thus are a great representation of stellar continuum emission and broad molecular absorption. The BP and RP spectra are each described by 55 coefficients of orthonormal Hermite functions, minimizing information loss during data compression \citep{2023A&A...674A...2D}. For the purpose of analysis, these coefficients are converted into wavelength-flux space using the {\tt GaiaXPy\footnote{\url{https://gaia-dpci.github.io/GaiaXPy-website/}}}  package \citep{2024zndo..11617977R}, resulting in 343 flux points spanning wavelengths from 336 nm to 1021 nm.

\subsection{The Training Sample} \label{sec:training_sample}

To classify XP spectra using neural networks, four categories of training samples are constructed: RSGs, oxygen-rich asymptotic giant branch stars (OAGBs), carbon-rich asymptotic giant branch stars (CAGBs), and miscellaneous sources (Misc.).

\begin{figure*}[htp]
	\centering
     \includegraphics[width=\textwidth]{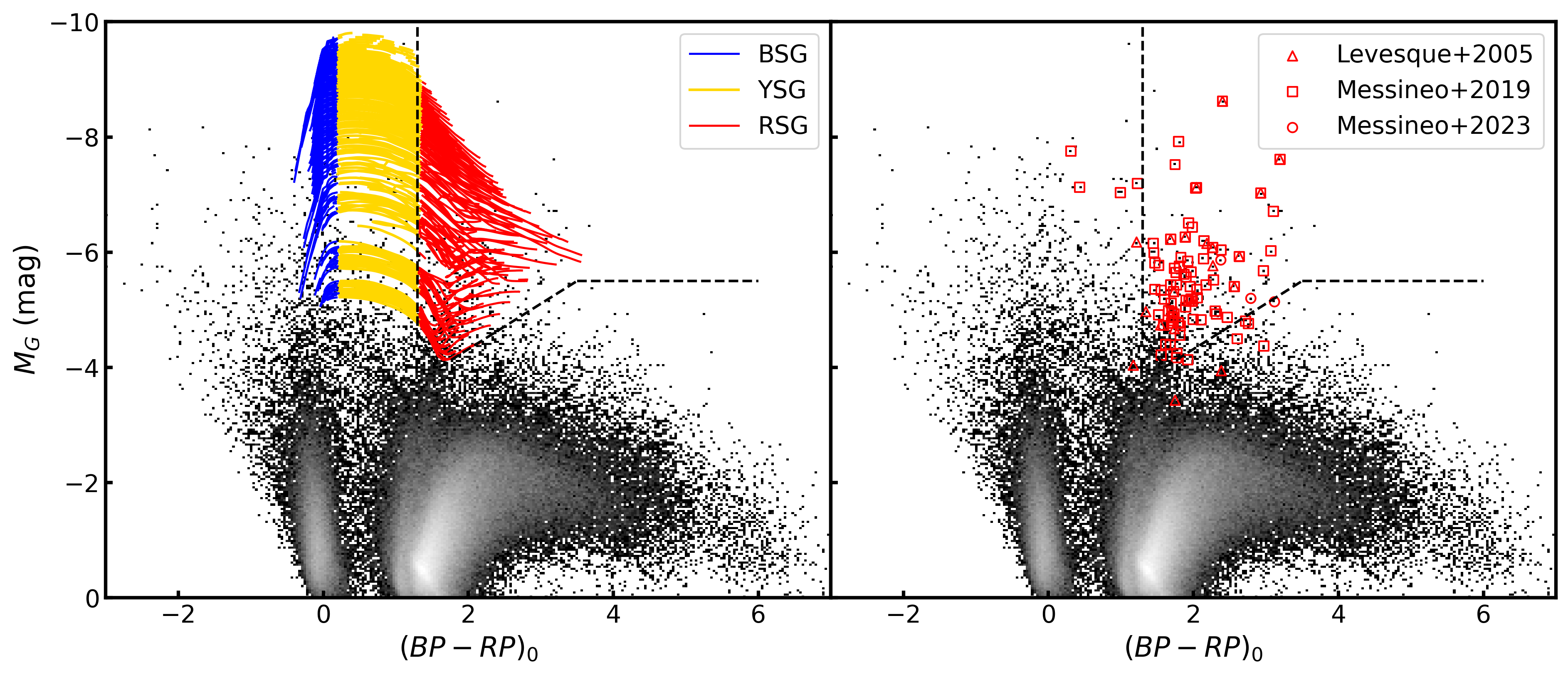}
	\caption{The Gaia CMD within 2 kpc. All the stars are marked by density, and the black dotted line represents the boundary of RSGs. The blue, yellow and red solid lines in the left panel represent the stellar evolutionary tracks of the MIST model, which are used to distinguish BSGs, YSGs and RSGs. Open triangles \citep{2005ApJ...628..973L}, squares \citep{2019AJ....158...20M} and circles \citep{2023A&A...671A.148M} on the right panel represent the known Galactic RSGs from the literatures. \label{fig:cmd_2kpc}}
\end{figure*}

The RSG seed sample is selected from a Gaia CMD of stars within a distance of 2 kpc. This is because the extinction and distance within this range are determined with high confidence, and thus the resulting CMD is reliable. Specifically, bright sources with available XP spectra are selected from the Gaia main catalog, requiring relative error of parallax smaller than 20\% and satisfying \texttt{phot\_g\_mean\_mag-5*log10(1000/parallax)+5<0}, which corresponds to absolute $G$ magnitudes without extinction correction brighter than 0 mag. This criterion is sufficiently inclusive, since the faint end of RSGs reaches about $G = -$4 mag \citep{2019A&A...629A..91Y, 2021A&A...646A.141Y}. To correct for extinction, we adopt reddening estimates from the DECaPS \citep{2025ApJ...992...39Z} and Bayestar19 \citep{2019ApJ...887...93G} maps with {\tt dustmaps} package \citep{2018JOSS....3..695G}, covering regions with $|b\vert < 10^{\circ}$ and the whole northern sky, respectively. If a star has the values measured by both of these maps, then the average is taken. Using the extinction law from \citet{2019ApJ...877..116W}, we derive $E(BP-RP) = 1.295 \times E(B-V)$ and $A_G = 2.446\times  E(B-V)$. The resulting CMD of bright sources within 2 kpc (Figure \ref{fig:cmd_2kpc}) contains $\sim$189,000 stars. Following \citet{2019A&A...629A..91Y, 2021A&A...646A.141Y}, we generate Modules for Experiments in Stellar Astrophysics (MESA; \citealt{2011ApJS..192....3P, 2013ApJS..208....4P, 2015ApJS..220...15P}) Isochrones \& Stellar Tracks (MIST; \citealt{2016ApJ...823..102C, 2016ApJS..222....8D}) evolutionary tracks for 8$-$40 $M_{\odot}$ stars with metallicities [Fe/H] = $-$0.2 to 0.2, including both rotating and non-rotating models. By selecting the core helium-burning phase, we distinguish between blue, yellow, and red supergiants (BSGs, YSGs, RSGs; see the left panel of Figure \ref{fig:cmd_2kpc}) by the effective temperatures ($T_{\mathrm{eff}}$) of the model (i.e., BSGs with $T_{\mathrm{eff}} > 7500\ \mathrm{K}$, YSGs with $4500 \ \mathrm{K} < T_{\mathrm{eff}} \leq 7500\ \mathrm{K}$ and RSGs with $T_{\mathrm{eff}} \leq  4500\ \mathrm{K}$), thereby defining empirical boundaries of the RSG region in the CMD. To verify the reliability of these boundaries, our RSGs sample is cross-matched with known Galactic RSGs identified by \citet{2005ApJ...628..973L}, \citet{2019AJ....158...20M} and \citet{2023A&A...671A.148M}, shown in the right panel of Figure \ref{fig:cmd_2kpc}. The majority of the known RSGs fall within our boundaries, confirming the reliability of this boundary. After removing sources not classified as K- or M-type stars and excluding carbon stars based on the SIMBAD classification, we obtain a final sample of 405 RSG candidates within 2 kpc.

The OAGB sample is compiled from the extreme OAGB catalog of \citet{2021ApJS..256...43S}, supplemented by OAGB stars selected by \citet{2025A&A...694A.152Z} with the criterion $K - W3 > 0.5$, which are OAGBs with Gaia RVS and missed by \citet{2021ApJS..256...43S}. The CAGB sample is also taken from the \citet{2021ApJS..256...43S} catalog. Both samples are required to have available XP spectra, resulting in 14,107 OAGB and 5,433 CAGB stars, respectively. The Misc. sample is drawn from the same set of bright sources within 2 kpc after excluding all stars in the RSG, OAGB, and CAGB samples, containing 186,352 objects, which is greatly larger than the RSGs sample.

\subsection{Data Pre-processing} \label{sec:pre-processing}
\subsubsection{Creating Extinction-Varied RSG Spectra}

To enlarge the RSG training sample, we add extinction to the 405 RSG spectra described above, generating artificially reddened spectra with extinction values of $A_V = 1,2,3,...,10$ mag with a step of 1 mag. In practice, each observed spectrum is de-reddened and re-reddened using the \citet{1999PASP..111...63F} extinction law (hereafter F99), assuming a standard Galactic extinction parameter of $R_V = 3.1$. The $A_V$ values for each star are taken from the DECaPS or Bayestar19 dustmaps, as mentioned in Section \ref{sec:training_sample}. This process is accomplished with the {\tt dust\_extinction} package \citep{2024JOSS....9.7023G}. The intrinsic (de-reddened) flux $F_{\rm int}(\lambda)$ is derived from the observed flux $F_{\rm obs}(\lambda)$ as
\begin{equation}
F_{\rm int}(\lambda) = \frac{F_{\rm obs}(\lambda)}{10^{-0.4{A_{\lambda}}}},
\end{equation}
with $A_{\lambda} = A_{V} k(\lambda)/R_V$ and $k(\lambda)$ being the reddening coefficient from the F99 model. A set of synthetic spectra is then produced by reapplying extinction of $A_V = 1-10$ mag to the intrinsic spectrum, which enhance the diversity of the training sample of RSGs. 

\subsubsection{Weighting Scheme and Normalization}

Class weights are introduced to address the imbalance caused by the significant differences in star counts across the four classes. The weight for each class $i$ is defined as
\begin{equation}
w_i = \frac{N}{C\times n_{i}},
\end{equation}
where $N$ is the total number of all spectra, $C=4$ is the number of classes, and $n_i$ is the number of stars in class $i$. This weighting scheme assigns higher importance to underrepresented classes, thereby reducing the bias caused by class imbalance. The sample sizes and corresponding weights for all four classes are summarized in Table \ref{tab:class_weight}. Clearly, RSGs carry the highest weight.

Moreover, to enable the neural network to distinguish RSGs based on spectrum, all spectra are normalized to their maximum flux, scaling them to the range of 0$-$1. 

\section{The Model}   \label{sec:model}

\subsection{The Model Architecture} \label{sec:model_architecture}
A variety of machine-learning approaches have been applied to Gaia XP spectra in recent literatures. To name a few, \citet{2024A&A...691A..98K} used gradient-boosted decision trees to perform large-scale label transfer and regression of stellar properties, producing the SHBoost catalog for $\sim$217 million XP sources. Similarly, \citet{2024ApJ...974..138Z} employed an XGBoost model for XP-based spectrophotometric inference, deriving stellar intrinsic colors and dust reddenings in a data-driven manner. Convolutional neural networks (CNNs), on the other hand, are particularly effective at learning localized spectral patterns and have been applied to XP-based classification problems such as carbon-star identification \citep[e.g.,][]{2025A&A...697A.107Y}.

\begin{figure*}[htp]
	\centering
     \includegraphics[width=\textwidth]{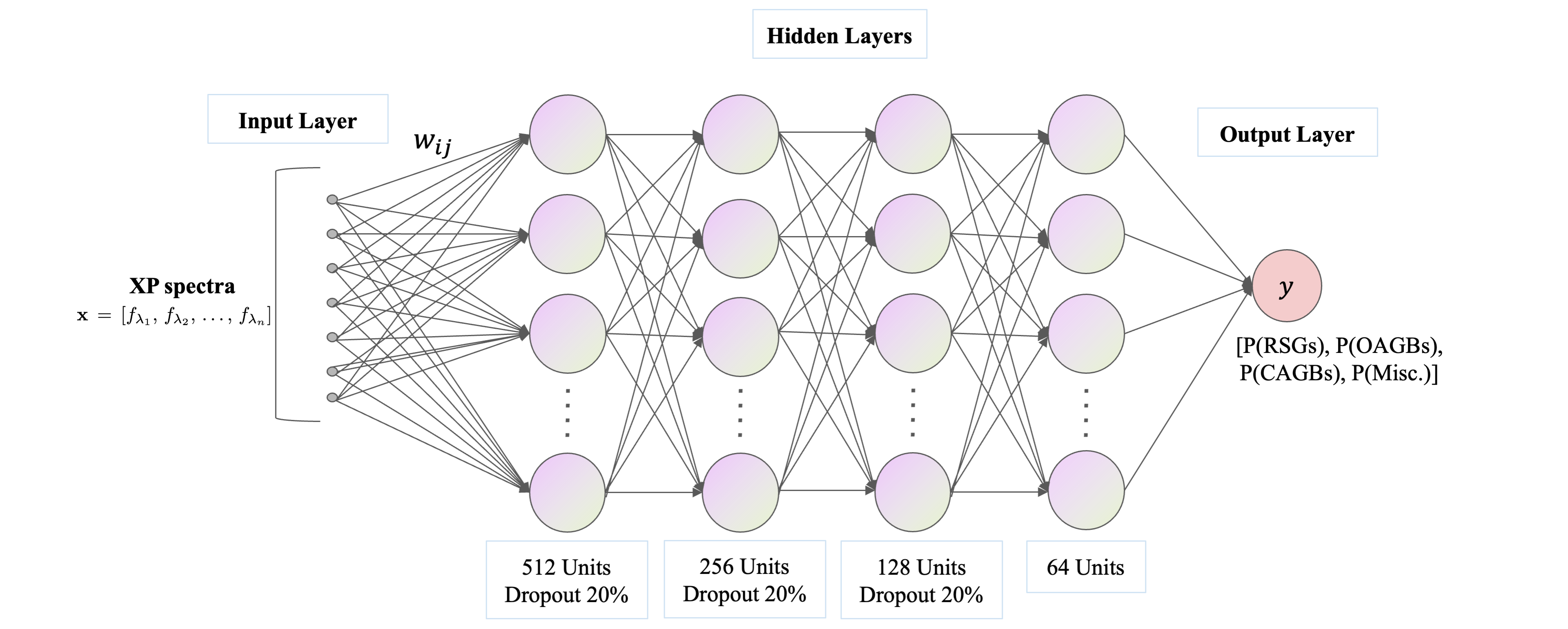}
	\caption{The overall architecture of the proposed network. This plot is modified from the original image contributed by srvmshr for ML Visuals\footnote{\url{https://github.com/dair-ai/ml-visuals}}.\label{fig:neural_network}}
\end{figure*}

To classify RSGs based on their Gaia XP spectra, we construct a multi-layer feedforward neural network (FNN) model. We adopt an FNN here because our inputs are fixed-length XP spectra and our goal is robust probabilistic classification with a simple, well-controlled architecture (i.e., strong baseline performance with minimal inductive assumptions compared to more specialized CNN designs). FNN is one of the most fundamental architectures in machine learning, capable of capturing complex nonlinear relations between input features and target classes. In the context of stellar spectral classification, FNN has proven effective in distinguishing flux variations and correlated spectral features across broad wavelength ranges \citep{1998MNRAS.298..361B, 1998MNRAS.295..312S, 2012A&A...538A..76N}. The architecture of the proposed FNN is shown in Figure \ref{fig:neural_network}. The input to the model is a one-dimensional normalized flux array representing each spectrum, denoted as $\mathbf{x} = [f_{\lambda_1},\, f_{\lambda_2},\, \dots,\, f_{\lambda_n}]$, where $n=343$ is the number of flux points spanning $336-1021$ nm. Each input vector is associated with a categorical label as $y \in \{ \mathrm{RSGs},\, \mathrm{OAGBs},\, \mathrm{CAGBs}, \,\mathrm{Misc.}\}$. The objective of the model is to approximate a mapping 
\begin{equation}
     f_{\theta} : \mathbf{x} \mapsto \mathbf{p}(y \mid \mathbf{x}),
\end{equation} 
where $f_{\theta}$ represents the neural network parameterized by weights $\theta$, and $\mathbf{p}(y \mid \mathbf{x})$ is the softmax probability distribution over the four stellar classes.

Our FNN consists of four fully-connected (dense) layers with Rectified Linear Units (ReLU) activation functions, batch normalization, and dropout regularization. Specifically, the first three layers contain 512, 256, and 128 neurons, respectively, each followed by batch normalization and a dropout rate of 0.2 to prevent overfitting \citep{2012arXiv1207.0580H, NEURIPS2018_36072923}. A fourth dense layer with 64 neurons refines higher-order representations, and the final output layer employs a softmax activation to produce class probabilities. The model is optimized using the Adam algorithm \citep{kingma2015adam} with a learning rate of $10^{-3}$. The loss function is the categorical cross-entropy, defined as
\begin{equation}
     \mathcal{L} = - \sum_{i=1}^{C} w_i \, y_{i} \, \ln \hat{y}_{i},
\end{equation}
where $y_{i}$ and $\hat{y}_{i}$ are the true and predicted probabilities for class $i$. The final output is a four-parameter probability vector $[\mathrm{P(RSG)}, \mathrm{P(OAGB)}, \mathrm{P(CAGB)}, \mathrm{P(Misc.)}]$, representing the probabilities that the star belongs to each class, with the sum four probabilities equal to 1.

\subsection{The Model Training and Validation} \label{sec:model_training}

The four sets of labeled samples are randomly divided into training and validation sets using ten random seeds, each following an 80\%-20\% ratio. For each run, the training set is used for parameter optimization, while the validation set provides an unbiased estimate of the model's generalization performance. These models are trained for up to 200 epochs with a batch size of 256. Two adaptive strategies are employed: (1) early stopping, which terminates training when the validation performance shows no improvement over 30 consecutive epochs, and (2) learning rate reduction, which halves the learning rate once the validation loss reaches a plateau. These measures ensure convergence stability and mitigate overfitting.

\begin{figure}[htp]
	\centering
     \includegraphics[width=0.5\textwidth]{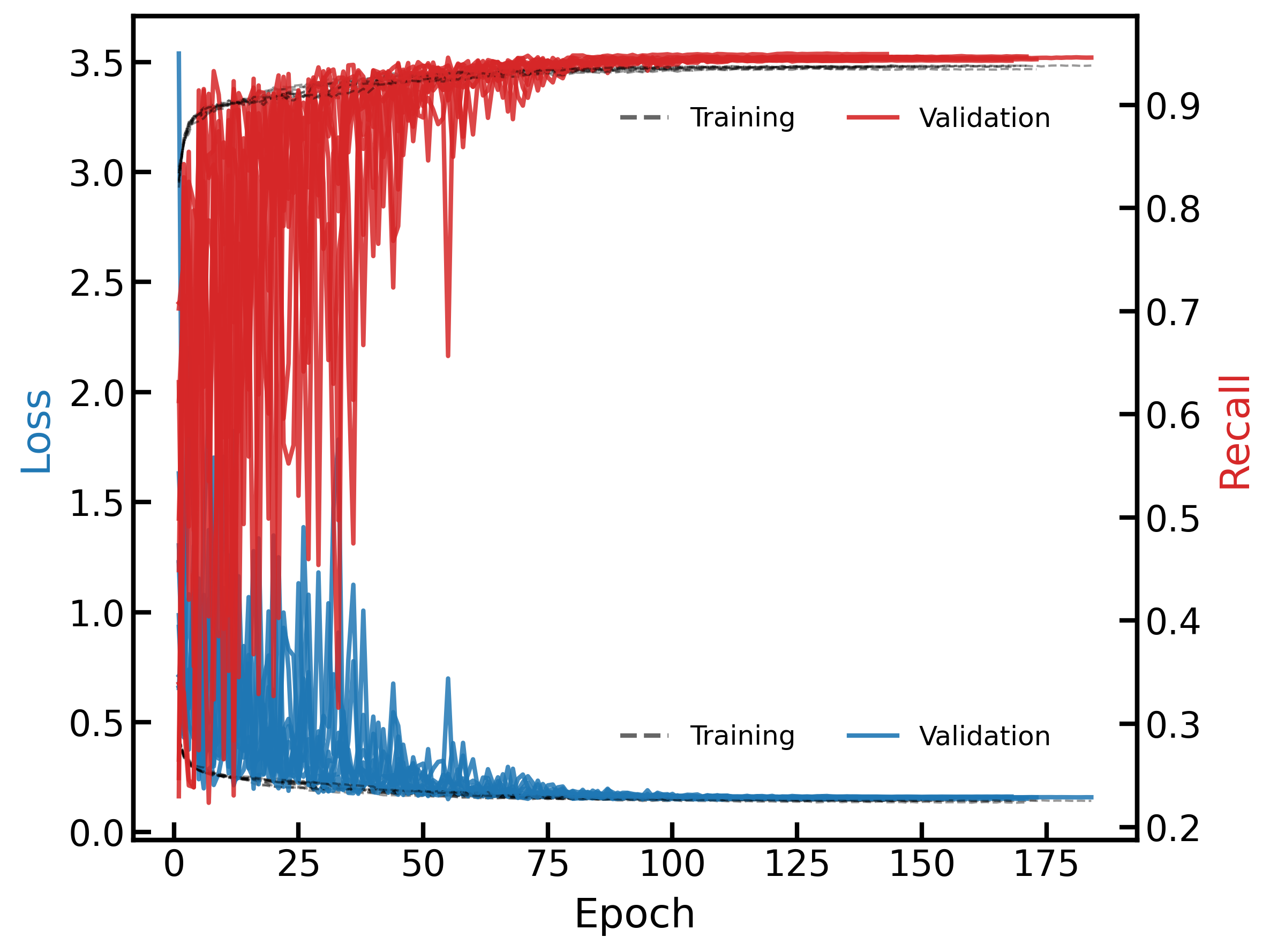}
	\caption{Evolution of the loss (blue, left axis) and recall (red, right axis) for the training (dashed lines) and validation (solid lines) sets as a function of epoch for ten independent runs. \label{fig:curve}}
\end{figure}

\begin{figure}[htp]
	\centering
     \includegraphics[width=0.5\textwidth]{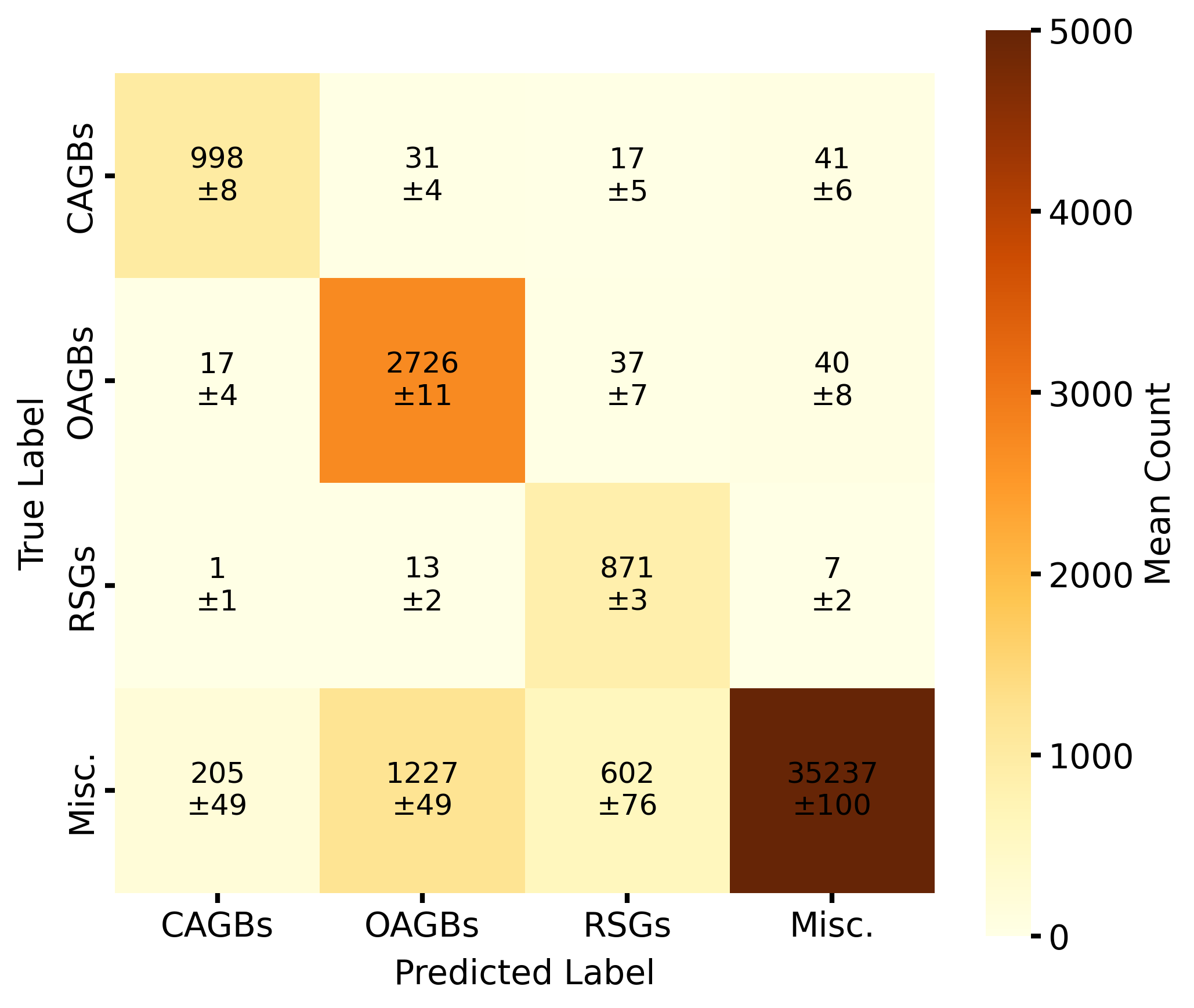}
	\caption{Mean confusion matrix of the validation set over ten runs. Each cell shows the average count and its 1$\sigma$ uncertainty. \label{fig:confusion_matrix}}
\end{figure}

\begin{figure}[htp]
	\centering
     \includegraphics[width=0.5\textwidth]{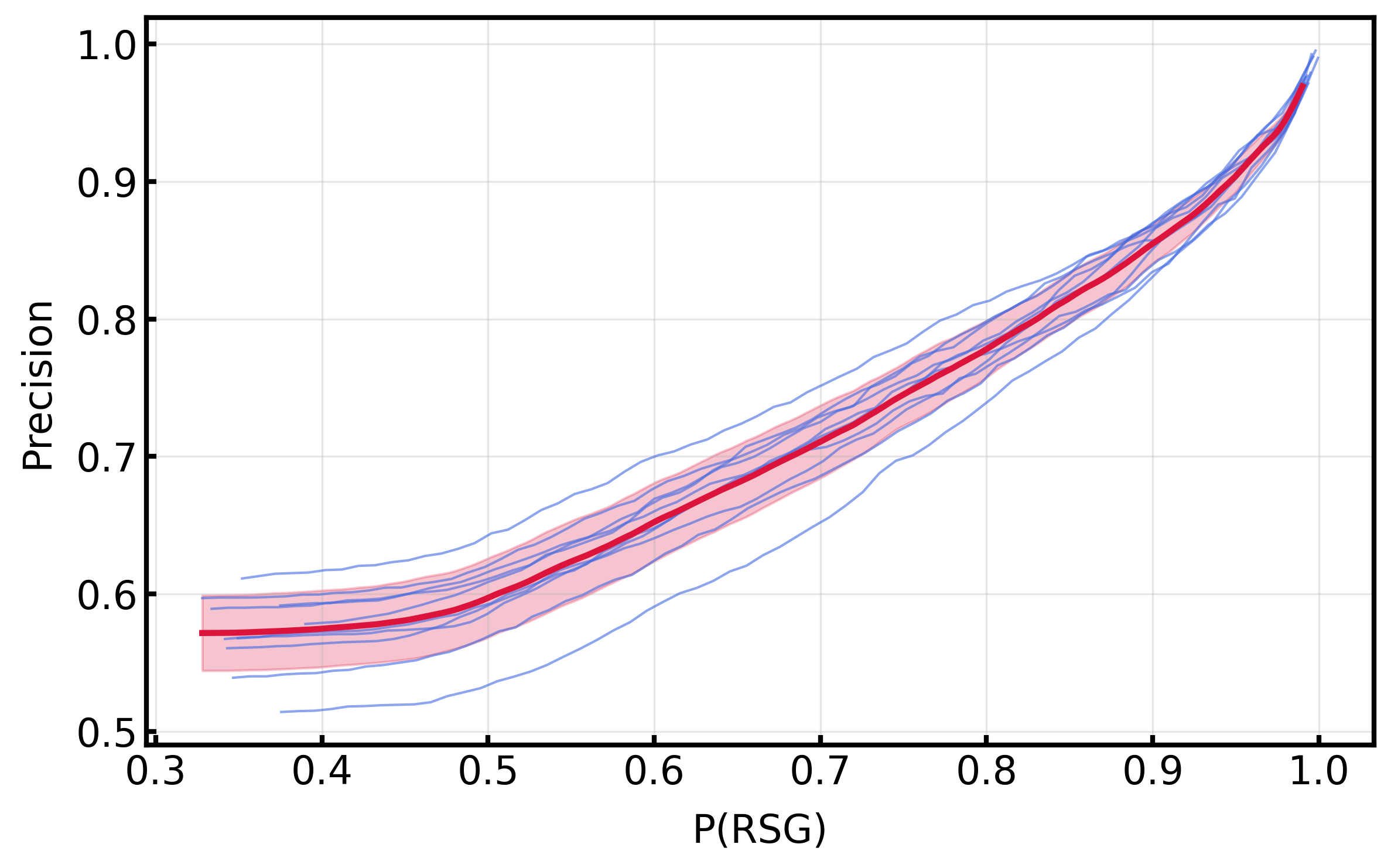}
	\caption{Precision as a function of $\mathrm{P(RSG)}$. Blue lines represent ten independent runs, and the red line and shaded region show the mean and 1$\sigma$ dispersion, respectively. \label{fig:precision}}
\end{figure}

The random seeds used in these ten runs and the number of training epochs are listed in Table \ref{tab:fnn_runs}. All of them achieved convergence within 200 epochs. Figure \ref{fig:curve} shows the evolution of the loss and recall for both the training and validation sets across all ten runs. Despite some fluctuations at early epochs, the curves converge consistently after approximately 75 epochs, and the final recall values exceed 90\% on average. Predictions are obtained by assigning each spectrum to the class with the highest probability, and compared with the true labels to compute the mean confusion matrix over the ten runs, as shown in Figure \ref{fig:confusion_matrix}. Each cell in the matrix reports the mean number with its 1$\sigma$ uncertainty. According to Table \ref{tab:fnn_runs}, the network achieved an average recall of 97.71\% and a precision\footnote{Precision is defined as the fraction of true RSGs among all sources classified as RSGs.} of 57.16\% for RSG classification, corresponding to an $f1$-score of 0.72.  As illustrated in Figure \ref{fig:confusion_matrix}, a certain level of contamination is present in the predicted RSGs. Because the model outputs the probability that each spectrum belongs to every class, we further examine how the precision varies with different thresholds of $\mathrm{P(RSG)}$. The corresponding relation is shown in Figure \ref{fig:precision}, where the blue curves represent the results of the ten independent runs, and the red line with shaded region denotes their mean and 1$\sigma$ dispersion. The precision increases monotonically with increasing $\mathrm{P(RSG)}$, indicating that contaminants generally exhibit relatively low $\mathrm{P(RSG)}$ values, even if they are the most probable class. By adopting a sufficiently high threshold, i.e., $\mathrm{P(RSG)}\geq 0.9$, a highly reliable RSGs sample can be obtained, with an average precision of 85.5\%.

\section{Result} \label{sec:result}
\subsection{The High-Confidence Galactic RSGs Sample}

For each run, the trained FNN model is applied to all 2,871,329 stars with available XP spectra in the $G < 12$ mag range to identify RSGs, and stars with $\mathrm{P(RSG)} \geq 0.9$ are selected as high-confidence RSG candidates, resulting in ten independent catalogs. The number of predicted RSGs and high-confidence RSGs are listed in Table \ref{tab:fnn_runs}. To balance completeness and purity, a star is regarded as a true RSG only if it appears in at least eight out of these ten catalogs. 

Furthermore, the following criteria are applied to ensure that these RSGs are located within the Milky Way:
\begin{enumerate}
\item A stellar probability greater than 99\% ({\tt classprob\_dsc\_combmod\_star}$>0.99$);
\item Stars not located in the LMC, i.e. the stars within the region $64^\circ<{\rm R.A.}<98^\circ$, $-78^\circ<{\rm Decl.}<-59^\circ$ and RV greater than 100 km/s, are excluded;
\item Stars not located in the SMC, i.e. the stars within the region $2^\circ<{\rm R.A.}<26^\circ$, $-76^\circ<{\rm Decl.}<-69^\circ$ and RV greater than 100 km/s, are excluded;
\end{enumerate}
The criteria are adopted because the SMC typically exhibits radial velocities of $\gtrsim 100~\mathrm{km~s^{-1}}$ \citep{2019A&A...629A..91Y}, the LMC generally has $\mathrm{RV} \gtrsim 166.5~\mathrm{km~s^{-1}}$ \citep{2021A&A...646A.141Y}, and the Magellanic Bridge connecting them is reported to have $\mathrm{RV} \sim 110~\mathrm{km~s^{-1}}$ \citep{2020A&A...641A.134S}. Therefore, applying a uniform cut of $\mathrm{RV} > 100~\mathrm{km~s^{-1}}$ provides an effective way to remove sources associated with the Magellanic system in this sky region. After applying these criteria, we obtain a final catalog containing 2,436 RSGs, listed in Table \ref{tab:RSGs_sample}. This catalog represents the most extensive and statistically robust compilation of Galactic RSGs to date. 

\begin{figure}[htp]
	\centering
     \includegraphics[width=0.5\textwidth]{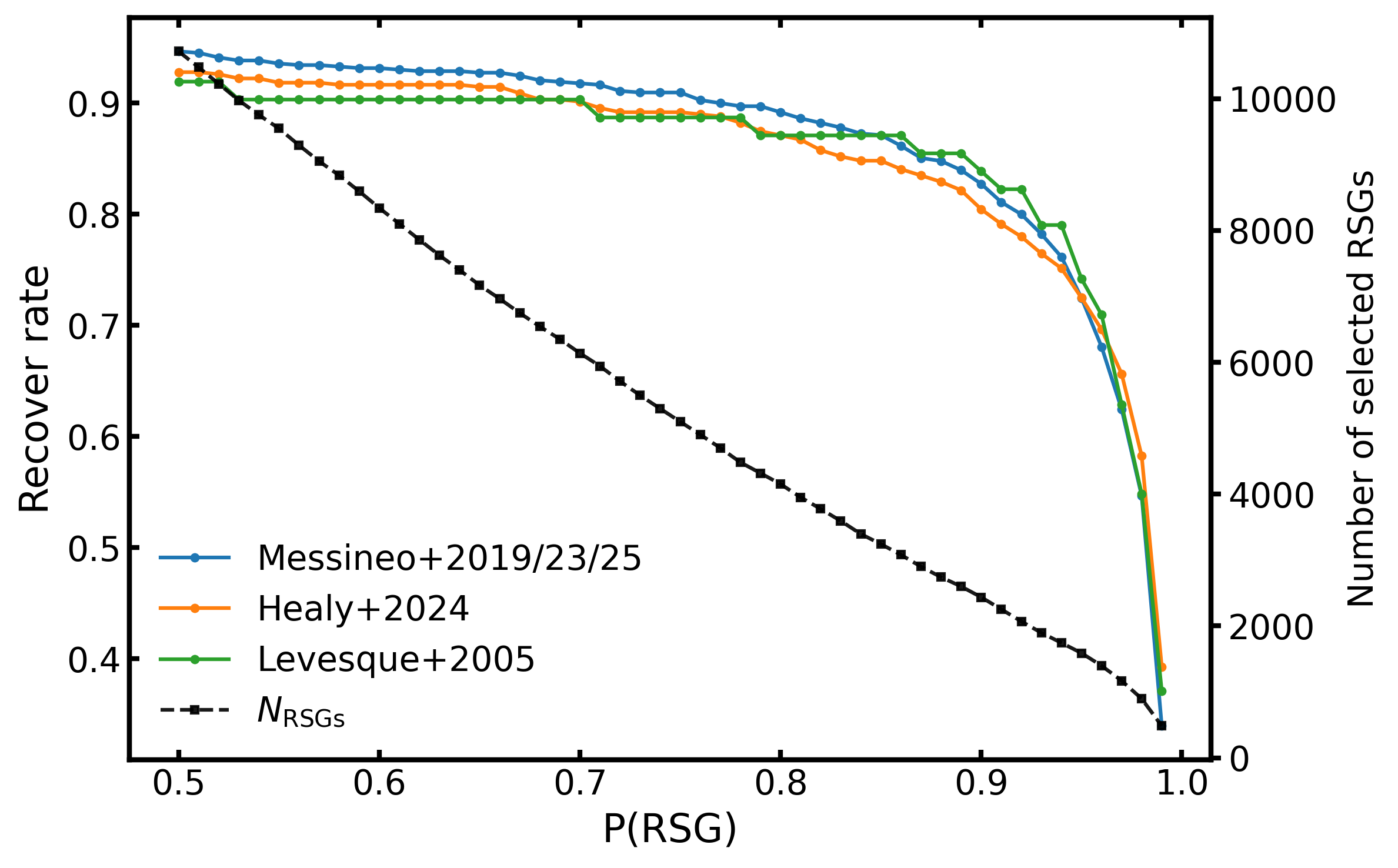}
	\caption{Recovery rate and sample size as a function of $\mathrm{P(RSG)}$. The left y-axis shows the recovery rate, plotted for three reference catalogs: \citet{2019AJ....158...20M} and \citet{2023A&A...671A.148M, 2025A&A...698A.282M} (blue solid line), \citet{2024MNRAS.529.3630H} (orange solid line), and \citet{2005ApJ...628..973L} (green solid line). The right y-axis shows the number of selected RSG candidates, $N_\mathrm{RSGs}$, plotted as a black dashed line with square markers.} \label{fig:sensitivity}
\end{figure}

To further assess the sensitivity of the adopted threshold and the reliability of our selection, we cross-match our sample with previously published Galactic RSG candidates and compute the recovery rate. We define the recovery rate as the ratio between (i) the number of sources in common between a given reference catalog and our selected sample and (ii) the number of sources in common between the same reference catalog and the initial sample used for our classification. We adopt the samples reported by \citet{2005ApJ...628..973L}, \citet{2019AJ....158...20M} and \citet{2023A&A...671A.148M, 2025A&A...698A.282M}, and \citet{2024MNRAS.529.3630H} as references, and these catalogs share 62 (among 74), 730 (among 762), and 525 (among 578) sources with our initial sample, respectively. The results are shown in Figure \ref{fig:sensitivity}. As $\mathrm{P(RSG)}$ increases, the number of our RSGs sample denoted by the black line decreases monotonically (right axis), whereas the recovery rates for the three reference catalogs denoted by the colorful lines remain high and nearly flat at first, and then drop sharply for $\mathrm{P(RSG)}>0.9$ (left axis). This behaviour indicates that sources with low $\mathrm{P(RSG)}$ are rarely present in the reference catalogs, and excluding them is therefore justified. Meanwhile, the reference catalog predominantly have $\mathrm{P(RSG)}>0.9$, as evidenced by the rapid decline in recovery rate when the threshold is higher than this value. This reinforces our choice of $\mathrm{P(RSG)}=0.9$ as a practical threshold: at this value, we recover the largest fraction of the reference catalogs with the smallest retained sample size (i.e., trading completeness for higher purity), achieving recovery rates above 80\% for all three reference catalogs.

We find no sources in common with \citet{2025MNRAS.538..101Z}, because their targets are substantially more distant and thus typically have $G>12$, falling outside our initial sample. The cross-match with \citet{2025A&A...694A.152Z} also yields only a small number of common sources. This may partly reflect unavoidable contamination in RVS-based selections; additionally, the sample of \citet{2025A&A...694A.152Z} mainly consist of late-type (later than M2) RSG candidates, a population that is under-represented in currently available Galactic samples. Among the 74, 365, and 578 RSG candidates in \citet{2005ApJ...628..973L}, \citet{2019AJ....158...20M}, and \citet{2024MNRAS.529.3630H}, only 27, 77, and 139 are later than M2, corresponding to an average fraction of $\sim 1/4$. Therefore, a high recovery with the above three reference catalogs naturally implies that only a limited overlap with \citet{2025A&A...694A.152Z} can be expected.

\begin{figure}[htp]
	\centering
     \includegraphics[width=0.5\textwidth]{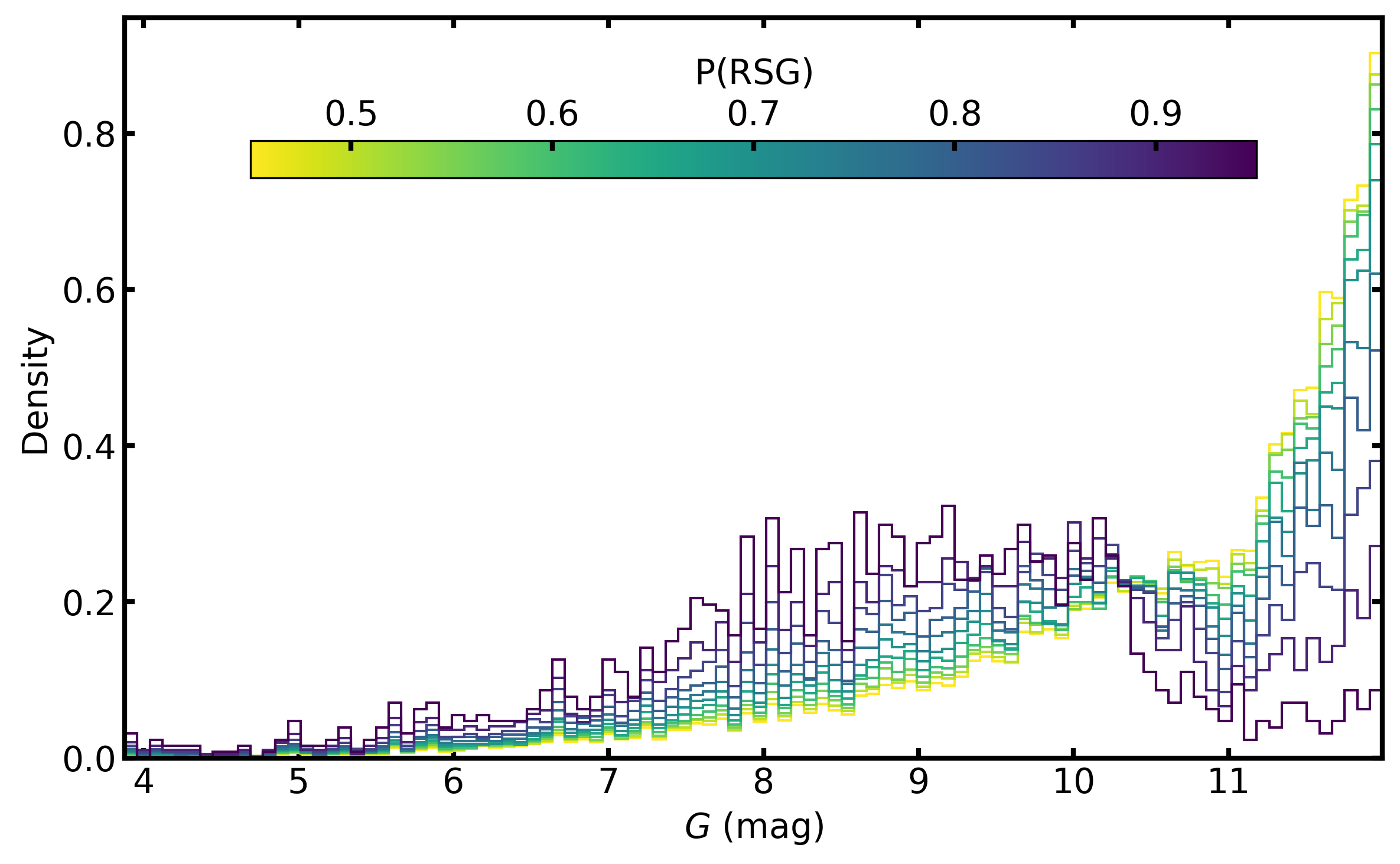}
	\caption{Normalized $G$-band magnitude distributions of the selected RSG samples with different $\mathrm{P(RSG)}$.} \label{fig:G_distribution}
\end{figure}

The $G$-band magnitude distributions of the selected RSG samples under different $\mathrm{P(RSG)}$ are shown in Figure \ref{fig:G_distribution}. A clear trend is that the relative contribution of faint candidates decreases as the $\mathrm{P(RSG)}$ is raised. This behaviour is expected because, at fainter $G$, Gaia XP spectra have lower signal-to-noise ratios and are more strongly affected by extinction and background systematics, which reduces the information available for separating RSGs from evolved contaminants (AGBs and luminous red giants). Consequently, a larger fraction of faint sources are assigned intermediate or low $\mathrm{P(RSG)}$ and are preferentially removed when adopting a stricter probability threshold.

\begin{figure*}[htp]
	\centering
     \includegraphics[width=\textwidth]{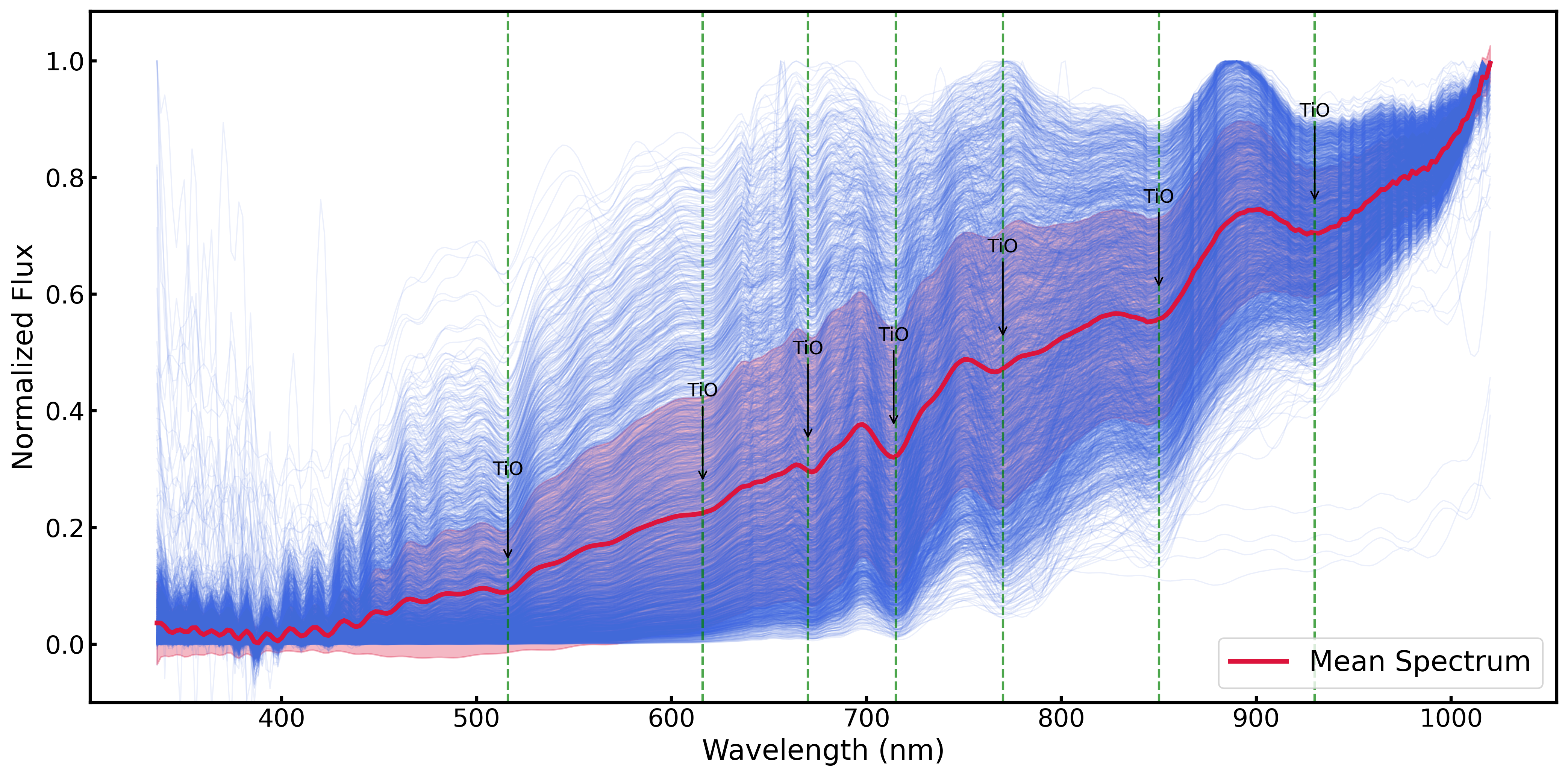}
	\caption{Stacked Gaia XP spectra of the newly identified RSGs by this work. Individual normalized spectra are shown in light blue, and their mean spectrum is indicated by the red line. Prominent TiO molecular absorption bands are marked with vertical dashed lines. \label{fig:RSG_spectra}}
\end{figure*}

The XP spectra of the newly identified RSGs are shown in Figure \ref{fig:RSG_spectra}. The spectra display a steadily rising continuum, reflecting their low effective temperatures and strong molecular opacity. Their mean spectrum shows several TiO bandheads between 500 and 1000 nm, characteristic of cool supergiant atmospheres with temperatures around $3500-4000\ \mathrm{K}$. These spectral features are consistent with the expectation that RSGs are oxygen-rich late-type stars.

\subsection{Spatial Distribution}

Both OB stars and RSGs represent massive stars and are generally believed to originate from the same star formation events, i.e., forming within the same molecular cloud, or young stellar cluster/OB association \citep{1994PASP..106...25G}. Consequently, they are expected to exhibit similar spatial distributions \citep{2016A&A...586A..46C}. To examine this, we investigate the spatial correlation between RSGs and OB stars. 

The OB star sample is adopted from \citet{2019MNRAS.487.1400C}, who identified 6,858 OB stars from the VST Photometric H$\alpha$ Survey Data Release 2 \citep{2014MNRAS.440.2036D} and the Gaia DR2. Combined with additional sources compiled from the literatures (the GOSC catalog from \citet{2013msao.confE.198M}, the OB star catalog selected from \citet{2014yCat....1.2023S}, and the catalog from the LAMOST Spectroscopic Survey of the Galactic Anti-centre (\citealt{2014IAUS..298..310L, 2015MNRAS.448..855Y})), the final sample comprises 14,880 OB stars. We cross-match these OB stars with Gaia DR3 to obtain updated and more accurate astrometric parameters, and computed their Galactocentric coordinates $(X, Y, Z)$ after selecting sources with $\sigma_{\varpi}/\varpi < 0.2$.

For RSGs, however, their parallaxes can be significantly biased due to their circumstellar dust envelopes, because the variability of the photocenter of RSGs brings uncertainty in the Gaia parallax \citep{2011A&A...528A.120C}, which is hard to measure. Therefore, for our RSG sample, we adopt the photogeometric distances derived by \citet{2021AJ....161..147B}, which incorporate a Galactic three-dimensional prior model to provide more reliable distance estimates. Five RSGs lack valid photogeometric distances, and for these objects the distances are calculated from the inverse of the parallax. These distances are then used to calculate the corresponding Galactocentric $(X, Y, Z)$ coordinates for the RSGs.

\begin{figure}[htbp]
    \centering
    \includegraphics[width=0.5\textwidth]{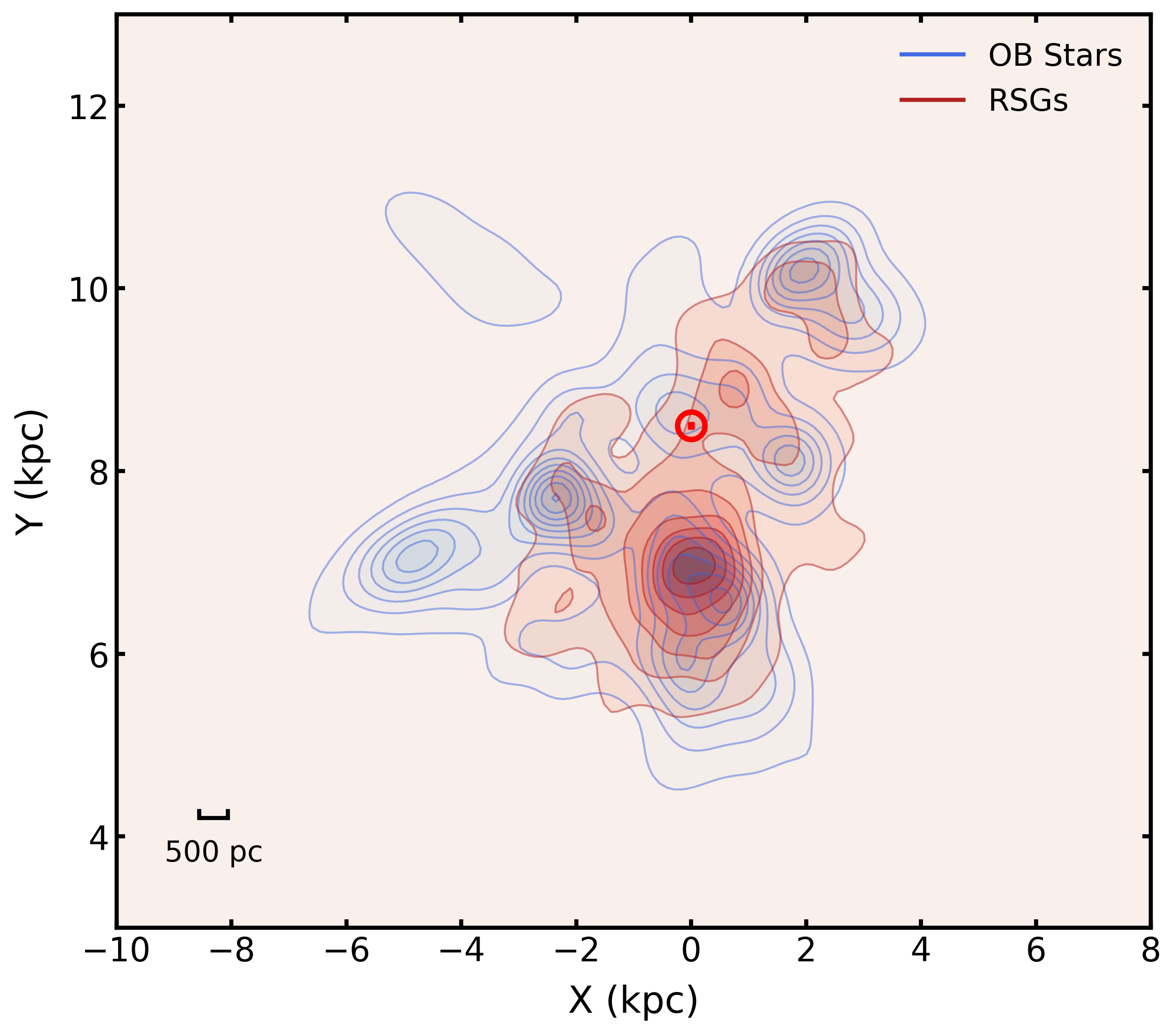}
    \caption{Spatial density distribution of RSGs (red contours) and OB stars (blue contours) in the Galactocentric $(X, Y)$ plane. The Sun is marked by a red circle at $(X, Y) = (0, 8.5)$ kpc. The scale bar of 500 pc is added in the lower-left corner for better context.}\label{fig:OB_stars}
\end{figure}

Figure \ref{fig:OB_stars} compares the projected spatial distributions of RSGs and OB stars in a bidimensional representation of the Galactic disk. Contours represent stellar number densities smoothed with a Gaussian kernel. Both populations show similar large-scale structures and concentrate along the Galactic spiral arms near the Solar neighborhood. The high-density regions of RSGs closely trace those of OB stars, indicating that they are spatially correlated and likely share a common origin in recent star-forming complexes. The slight displacement between some RSG and OB density peaks may arise from stellar evolution effects, as RSGs represent a later evolutionary stage of massive stars and could have moved slightly from their birthplaces due to stellar motions. As noted by \citet{2020A&A...644A..62C}, RSGs could be expected to move $\sim 100$ pc in 20 Myr, which is much larger than the typical size of an OB association. 

\begin{figure*}[htbp]  
    \centering
    \includegraphics[width=\textwidth]{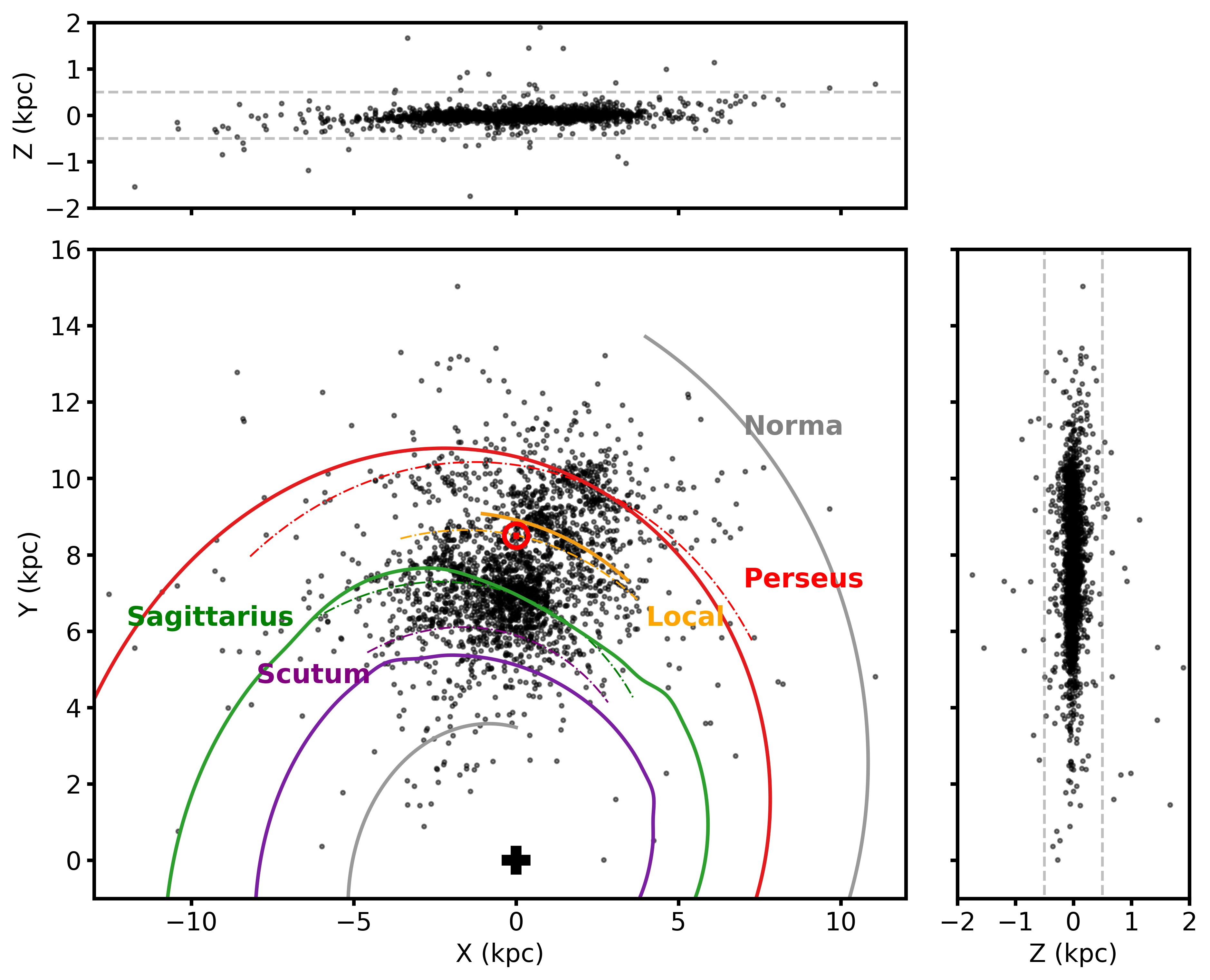}
    \caption{Three-dimensional spatial distribution of the identified RSGs in the Galactocentric coordinate system. In the $(X, Y)$ plane, the colorful solid curves indicate the spiral-arm loci from \citet{2024RNAAS...8...17O}, while the corresponding dashed curves show the arm parameterizations of \citet{2019MNRAS.487.1400C} (Scutum: purple; Sagittarius: green; Local: orange; Perseus: red). The Norma arm is only present in \citet{2024RNAAS...8...17O} and is marked with a gray solid line. The Sun is denoted by the red circle and the Galactic center by the black cross. The gray dashed lines in the upper and right panels mark $Z = \pm 0.5$ kpc.}\label{fig:RSG_spatial}
\end{figure*}

The three-dimensional distribution of the identified RSGs is further illustrated in Figure \ref{fig:RSG_spatial}. The projection in the $(X, Y)$ plane reveals that the majority of RSGs are distributed along several well-defined Galactic spiral arms, following the parameterizations of \citet{2019MNRAS.487.1400C} and \citet{2024RNAAS...8...17O}. The former is a spiral-arm model based on OB stars comprising four arms (dashed curves), whereas the latter is a pulsar-derived free electron density model that includes an additional Norma arm (solid curves). This trend is consistent with what we saw in \citet{2018MNRAS.475.2003D, 2019AJ....158...20M, 2024MNRAS.529.3630H, 2025A&A...698A.282M}, but covers a much larger spatial range. Visually, the Perseus arm is the most prominent, consistent with the overdensity traced by the green dots in Figure 11 of \citet{2025A&A...698A.282M}. Our RSG candidates associated with the Local arm are more numerous than in previous literatures, and their locus appears more consistent with the arm of \citet{2024RNAAS...8...17O}. The Sagittarius arm also intersects the high-density region toward the Galactic-centre direction. In particular, the \citet{2024RNAAS...8...17O} model follows a population of RSG candidates on the far-side arm segment at $-6 < X < -2$ kpc. The structure around $(X,Y)\sim (-3,\,7.5)$ kpc is also consistent with the cyan dots reported by \citet{2025A&A...698A.282M}. For the Scutum arm, however, the \citet{2019MNRAS.487.1400C} and \citet{2024RNAAS...8...17O} parameterizations differ by $\sim$1 kpc, likely reflecting the increasing difficulty of distance and extinction determinations toward the inner Galaxy. As a result, it is challenging to unambiguously assign the corresponding structures to a specific arm in this region. The overdensity traced by the orange points in Figure 11 of \citet{2025A&A...698A.282M} is clearly present in our sample at $-5 < X < -1$ kpc and $5 < Y < 7$ kpc. However, in the \citet{2024RNAAS...8...17O} model this structure could lie on the inner Scutum arm or in the inter-arm region between Scutum and Sagittarius, whereas in the \citet{2019MNRAS.487.1400C} model it appears to cross the Scutum arm. A subset of RSG candidates near $Y\sim3$ kpc may be related to the Norma arm, although this association remains uncertain. In addition, several filaments features, particularly those at $-5 < X < 0$ kpc and $9 < Y < 10$ kpc, are clearly visible in our distribution and are in good agreement with the structures reported by \citet{2025A&A...698A.282M}, which are also shown in Figure 6 of \citet{2024MNRAS.529.3630H}.

As for the vertical distributions in the $(X, Z)$ and $(Y, Z)$ planes, the vast majority ($\sim98\%$) of RSGs are confined within $\vert Z \vert < 0.5$ kpc, indicating that they are mainly distributed in the thin disk. This is consistent with their massive-star origin and short evolutionary timescales, which restrict them to the Galactic plane.

The overall spatial distribution of the identified RSGs aligns well with the expected locations of the Galactic spiral arms, confirming the reliability of our classification. Moreover, the clear correspondence between RSGs and the spiral arm features reinforces their effectiveness as robust tracers of Galactic structure, particularly in regions where OB stars or H II regions are heavily obscured.

\section{Summary} \label{sec:summary}

RSGs, which are massive stars in the core-helium-burning phase of late stellar evolution, serve as excellent tracers of the Galactic disk's structure and kinematics. While the number of RSGs identified in extragalactic galaxies has grown substantially in recent years, the Galactic RSG population remains poorly characterized. In this work, we train a feedforward neural network classifier, which is capable of assigning a probability for each XP spectrum to belong to an RSG. Our data is randomly divided into training and validation subsets with an 80\%-20\% ratio, and the model is trained ten times. When adopting a probability threshold of $\mathrm{P(RSG)} \geq 0.9$ for high-confidence candidates, the classifier achieved an accuracy of 85.5\%. Each of the ten trained models is applied to all Gaia XP spectra of stars with $G < 12$ mag, yielding ten independent catalogs of high-confidence RSG candidates. To minimize the effects of random choices in the training and validation sets, only stars appearing in at least eight of these ten catalogs are considered true RSGs, resulting in a final sample of 2,436 Galactic RSGs. This catalog recovers over 80\% of the RSGs reported in previous studies, confirming its reliability.

Further investigation reveals a spatial correlation between RSGs and OB stars, consistent with the expectation that both populations originate from the same massive star-forming regions. The displacement between the density peaks of RSGs and OB stars might be due to the fact that RSGs, as a subsequent evolutionary stage, have slightly moved away from the OB associations. The RSGs are generally confined in $\vert Z \vert < 0.5$ kpc of the Galactic plane, indicating their location in the thin disk. These RSGs trace four major spiral arms in the Solar neighborhood clearly in the $(X, Y)$ projection, in agreement with expectations in literatures, but extending over a larger spatial range.

These results demonstrate that RSGs can serve as an independent and powerful tracer of the Milky Way's spiral structure, offering significant potential for future studies of Galactic morphology and evolution. Given the abundant observational data in the Milky Way, the RSGs sample obtained in this work also provides a valuable foundation for more detailed studies of their physical properties, thereby greatly enhancing our understanding of these massive stars.

\begin{acknowledgments}
We would like to thank the anonymous referee for the constructive suggestions that definitely improved this work. This work is supported by the National Natural Science Foundation of China (NSFC) through grants Nos. 125B2060, 12133002 and 12203025, and Shandong Provincial Natural Science Foundation through project ZR2022QA064. This work has made use of data from the European Space Agency (ESA) mission {\it Gaia} (\url{https://www.cosmos.esa.int/gaia}), processed by the {\it Gaia} Data Processing and Analysis Consortium (DPAC, \url{https://www.cosmos.esa.int/web/gaia/dpac/consortium}). Funding for the DPAC has been provided by national institutions, in particular the institutions participating in the {\it Gaia} Multilateral Agreement. This work has made use of the Python package GaiaXPy, developed and maintained by members of the Gaia Data Processing and Analysis Consortium (DPAC), and in particular, Coordination Unit 5 (CU5), and the Data Processing Centre located at the Institute of Astronomy, Cambridge, UK (DPCI).
\end{acknowledgments}

%

\vspace{5mm}


\software{astropy \citep{2013A&A...558A..33A,2018AJ....156..123A},
          TOPCAT \citep{2005ASPC..347...29T},
          tenserflow \citep{tensorflow2015-whitepaper},
          sklearn \citep{scikit-learn},
          GaiaXPy \citep{2024zndo..11617977R}}

\facility{Gaia}

\bibliography{Galactic_RSGs}{}
\bibliographystyle{aasjournalv7}

\begin{table}[htbp]
\centering
\caption{The number of stars and weights of the training samples.}\label{tab:class_weight}
\renewcommand{\arraystretch}{1.2}
\begin{tabular}{lcc}
\hline
\hline
Class & Number of Stars & Weight \\
\hline
RSGs  & 4455  & 11.8040 \\
OAGBs & 14107 & 3.7277 \\
CAGBs & 5433  & 9.6791 \\
Misc. & 186352 & 0.2822 \\
\hline
\end{tabular}
\end{table}

\begin{table*}[htbp]
    \centering
    \caption{Performance metrics of the ten FNN runs with different random seeds.}
    \label{tab:fnn_runs}
    \begin{tabular}{ccccccc}
    \hline
    \hline
    Random Seeds & Epochs & Precision$^a$ & Recall$^a$ & $f1$-score$^a$ & N(RSGs)$^b$ & N(RSGs with P $\geq0.9$)$^c$ \\
    \hline
     8183 & 173 & 59.17\% & 97.42\% & 0.7362 & 14170 & 3005 \\
     808  & 87  & 51.42\% & 97.76\% & 0.6739 & 17766 & 3237 \\
     6401 & 184 & 61.13\% & 97.42\% & 0.7512 & 13699 & 2795 \\
     8906 & 139 & 56.06\% & 97.64\% & 0.7122 & 14789 & 2917 \\
     3752 & 102 & 53.91\% & 97.53\% & 0.6944 & 14902 & 2731 \\
     5224 & 171 & 57.82\% & 97.08\% & 0.7248 & 15035 & 2997 \\
     8484 & 143 & 59.70\% & 98.09\% & 0.7423 & 14701 & 2873 \\
     1734 & 168 & 56.73\% & 97.98\% & 0.7185 & 14322 & 3182 \\
     8878 & 155 & 56.77\% & 98.32\% & 0.7198 & 14443 & 3137 \\
     4461 & 150 & 58.92\% & 97.87\% & 0.7356 & 13539 & 3227 \\
    \hline
    \end{tabular}

    \vspace{2mm}
    \begin{flushleft}
    \footnotesize
    \textit{Notes.} $^a$ Precision, recall and $f1$-score refer to the RSG class on the validation set for each run. \\
    $^b$ Number of stars with $G<12$ mag that are classified as RSGs by the model (i.e. $\mathrm{P(RSG)}$ is the largest among the four classes). \\
    $^c$ Number of high-confidence RSGs with $\mathrm{P(RSG)}\geq 0.9$ for each run.
    \end{flushleft}
\end{table*}

\begin{table*}[htbp]
\centering
\caption{The high-confidence Galactic RSGs catalog.}
\label{tab:RSGs_sample}
\renewcommand{\arraystretch}{1.1}
\begin{tabular}{lccccccc}
\hline
\hline
{\tt source\_id} & R.A. & Decl. & $\mathrm{P(RSG)}^{a}$ & $\mathrm{Distance_{p,geo}}$ & $X$ & $Y$ & $Z$ \\
                 & (deg) & (deg) & (\%) & (kpc) & (kpc) & (kpc) & (kpc) \\
\hline
528405467039323008  & 0.7169 & 66.3424 & 93.09 & 1.58 & 1.39 & 9.24 & 0.11 \\
432083335007646720  & 1.7855 & 64.4900 & 98.70 & 6.46 & 5.69 & 11.55 & 0.23 \\
429264908751546752  & 1.9240 & 60.1607 & 93.49 & 2.77 & 2.46 & 9.78 & -0.11 \\
431678852171577216  & 2.3597 & 63.9540 & 96.22 & 2.66 & 2.34 & 9.76 & -0.07 \\
429999760479433520  & 2.4015 & 52.6678 & 99.33 & 3.98 & 3.51 & 10.38 & 0.01 \\
431476232769606368  & 3.4960 & 53.7420 & 97.45 & 3.47 & 3.04 & 10.17 & 0.07 \\
528727761385369728  & 4.1765 & 57.5508 & 94.16 & 1.57 & 1.36 & 9.27 & 0.13 \\
428817510598195584  & 4.6099 & 60.9025 & 97.60 & 2.87 & 2.51 & 9.89 & -0.09 \\
428817128336340864  & 4.7763 & 60.9341 & 97.08 & 2.64 & 2.31 & 9.78 & -0.07 \\
428786311954846080  & 5.0051 & 60.6392 & 97.39 & 1.01 & 0.88 & 9.99 & -0.04 \\
430617754731938816  & 5.1316 & 52.8688 & 97.21 & 2.46 & 2.14 & 9.71 & -0.09 \\
430464235421496320  & 5.1815 & 51.8800 & 99.80 & 1.12 & 0.98 & 9.05 & -0.02 \\
430463582586531072  & 5.1871 & 51.8082 & 99.20 & 2.75 & 2.40 & 9.85 & -0.04 \\
... & ... & ... & ... & ... & ... & ... & ... \\
\hline
\end{tabular}

\vspace{2mm}
\begin{flushleft}
\footnotesize
$^{a}$ The value of $\mathrm{P(RSG)}$ for each star is calculated as the mean probability across the ten independent catalogs. \\
(This table is available in its entirety in machine-readable form.)
\end{flushleft}

\end{table*}

\end{CJK*}
\end{document}